\documentstyle[aps,pra]{revtex}
\begin{document}

\twocolumn[\hsize\textwidth\columnwidth\hsize\csname@twocolumnfalse\endcsname
\draft 

\title{Algebraic solution of master equations}
\author{R. Rangel, L. Carvalho}
\address{ Instituto de F\'\i sica, Universidade Federal do Rio de Janeiro,Caixa Postal 68528, 21945-970, Rio de Janeiro, Brazil.}
\maketitle
\begin{abstract}
We present a simple analytical method to solve master equations for finite temperatures and any initial conditions, which consists in the expansion of the density operator into normal modes. These modes and the expansion coefficients are obtained algebraically by using ladder superoperators. This algebraic technique is successful in cases in which the Liouville superoperator is quadratic in the creation and annihilation operators.
\end{abstract}

\pacs{}]

\vskip2pc

In general, any physical system is affected by its surroundings since, even in laboratory, it is hard to avoid its interaction with the environment. Furthermore the dissipative processes and the continuous fluctuations of the environment generate decoherence processes, which are responsible to an irreversible loss of information of the system. Thus the environment plays a fundamental role in the dynamics of any system so that in any real physical situation it is necessary to analyze its effects. 

In a Hamiltonian formalism, certain types of reservoir may be simulated by a bath of harmonic oscillators linearly coupled to the system and, in some cases, the solution of the full problem of coupled system-reservoir dynamics may be exactly obtained in the density operator formalism~\cite{collett}. Nevertheless, we are in general only interested in observing the system so that we should eliminate the environment by tracing over the bath variables, which provides us the reduced density operator of the system, $\rho(t).$ 

The approximated dynamics of the reduced density operator does not need of the full solution: we may generate, through the trace over the bath and suitable approximations, master equations that describe its evolution. These equations become irreversible due to the fact that the system-reservoir correlations are suppressed in the reduced density operator. The most common approximation consists in adopting the Born-Markov limit in which the system-bath coupling is weak and the memory of the reservoir is short. In this limit, the master equation becomes local and the effects of the reservoir are resumed in the form of a non-Hamiltonian Lindblad term. The general form of this type of equation is \cite{milburn,miguel}

\begin{equation}\label{eqm}
\frac{\partial \rho(t)} {\partial t} = \frac{1} {i \hbar}[H, \rho(t)] + {\cal L}\rho(t) \, ,
\end{equation}
where $H$ is the Hamiltonian of the system and ${\cal L}\rho$ is the Lindblad term generated in the elimination of the bath degrees of freedom. 

Several analytical methods to solve master equations of this type for any initial condition can be found in Refs.~\cite{nicim,sonia,gilles,romero,walls,perinova,englert}. The simple case of an oscillator in a zero-temperature reservoir has been solved by Guerra {\it et al.}~\cite{nicim}. Mokarzel~\cite{sonia} has solved the case of two coupled oscillators at zero temperature and in the presence of pump in both modes by using a method based on the Lie algebra for superoperators. This method consists in decomposing the exponential of the total Liouvillian into simpler exponentials. Gilles and Knight~\cite{gilles} have solved a master equation in which the Lindblad form contains terms non-quadratic in the creation and annihilation operators, like $a^2 \rho \hskip 0.05cm a^{\dagger \hskip 0.1cm 2}.$ A wide class of problems at zero temperature, including Hamiltonians and Lindblad terms non-quadratic in the operators $a$ and $a^{\dagger},$ has been solved by Klimov and Romero~\cite{romero}. In this case, superoperators belonging to a deformed algebra have been used. Walls and Milburn~\cite{walls} have used quasidistributions to obtain the solution of an oscillator coupled to a thermal reservoir without the usual rotating-wave approximation. Pe\u rinov\'a and Luk\u s~\cite{perinova} also have used quasidistributions to solve a master equation at finite temperature with a Hamiltonian of the form $a^{\dagger \hskip 0.1cm 2} a^{2}.$ Briegel and Englert~\cite{englert} have elaborated a method based on the expansion of the density operator of the problem into the eigenstates of the Liouvillian. The eigenstates have been obtained by using differential equations and, the expansion coefficients, by taking a scalar product of the initial density operator with a dual basis orthonormal to the eigenstate basis. This method has been used~\cite{englert} to solve some problems as a single oscillator in a finite temperature bath and a two-level system coupled to a cavity field mode through a Jaynes-Cummings type interaction at zero temperature.

In this paper we present an algebraic technique to obtain the solution of master equations, which consists in the expansion of the density operator into the eigenstates of the Liouvillian. The eigenstates and the expansion coefficients are easily obtained in an algebraic way: we define step-down superoperators, which are used for obtaining the steady state and the expansion coefficients, and step-up superoperators, which enable us to obtain the remaining eigenstates. In the next sections we present some examples where this technique is successful: harmonic oscillator in a zero-temperature bath and in a thermal bath, two-level system in a thermal bath and two coupled oscillators in a zero-temperature bath and in a finite temperature bath. To our knowledge the last case has not yet been solved.

\section{Damped harmonic oscillator}

Eq.~(\ref{eqm}) may be rewritten in a more compact form
 
\begin{equation}\label{eqmK}
\frac{\partial \rho(t) }{ \partial t} =  {\cal K}\rho(t)\, ,
\end{equation}
where the Liouvillian superoperator ${\cal K}$ is defined as

\begin{equation}\label{K.}
{\cal K} = \frac{1  }{ i \hbar} (H. - .H) + {\cal L} \hskip 0.3cm .
\end{equation}
Eq.~(\ref{eqmK}) has the formal solution,

\begin{equation}\label{formal}
\rho(t) = e^{\displaystyle{ {\cal K}t}} \rho(0) \, .
\end{equation}
However, it is not trivial, in general, to obtain a useful expression for the density operator. To allow the action of this time evolution operator we will expand the initial density operator into the eigenstates of the Liouvillian. 

In order to exemplify our method, we will analyze the behavior of a harmonic oscillator in contact with a zero-temperature reservoir. Then, we will solve a master equation of type (\ref{eqm}), where the Hamiltonian of the system is given by  

\begin{equation}\label{Ha}
H = \hbar \omega a^{\dagger} a
\end{equation} 
and the Lindblad superoperator that simulates the reservoir has the form \cite{milburn,miguel}

\begin{equation}\label{L}
{\cal L} = \frac{\gamma }{ 2} (2 a..a^{\dagger} - a^{\dagger} a. - .a^{\dagger} a) \, ,
\end{equation}
being $\gamma$ the damping rate of the system energy due to the action of the reservoir. 

Here we have adopted the notation $A.$ $(.A)$ for superoperators that represent the simple action of an operator $A$ to the left (right) on the target operator, $\rho,$ i.e., $A. \rho:= A \rho$ $(.A \rho := \rho A).$ In order to avoid confusion between operators and superoperators, the latter will be denoted by either a Calligraphic letter or a Roman letter followed or preceded by a point. It is easy to demonstrate that superoperators of this form obey the following relations:

\begin{eqnarray}\label{A.B.}
A.B. &=& AB. \hskip 0.2cm ,  \hskip 0.2cm .A.B = .BA \, , \nonumber \\
.BA. &=& A..B \hskip 0.2cm ,  \hskip 0.2cm .1 = 1. \equiv 1 \, ,
\end{eqnarray}
which, obviously, only have a meaning if acted on an arbitrary target operator. 

By defining the commutator between two superoperators in the usual form, $[{\cal A},{\cal B}] := {\cal A}{\cal B} - {\cal B}{\cal A},$ we can show that almost all the commutator properties among operators remain valid here, for example, 

\begin{eqnarray}\label{comutador}
\lbrack\lambda {\cal A}+ \delta {\cal B},{\cal C}\rbrack &=& \lambda \lbrack{\cal A},{\cal C}\rbrack + \delta \lbrack{\cal B},{\cal C}\rbrack \, , \nonumber \\
\lbrack{\cal B},{\cal A}\rbrack &=& - \lbrack{\cal A},{\cal B}\rbrack \, , \nonumber \\
\lbrack{\cal A}{\cal B},{\cal C}\rbrack &=& \lbrack{\cal A},{\cal C}\rbrack{\cal B} + {\cal A}\lbrack{\cal B},{\cal C}\rbrack \, ,
\end{eqnarray}
where $\lambda$ and $\delta$ are complex numbers. Furthermore, from relations (\ref{A.B.}), it is easy to obtain the following properties:       

\begin{eqnarray}\label{comutador.}
[A.,B.] &=& [A,B]. \hskip 0.2cm ,  \hskip 0.2cm \lbrack .A,.B \rbrack = .\lbrack B,A \rbrack \, , \nonumber \\
\lbrack A.,.B \rbrack &=&0 \hskip 0.3cm .
\end{eqnarray}

Using the above relations and the definition of ${\cal K},$ we may obtain the commutation relations

\begin{eqnarray}\label{comutadora}
\lbrack {\cal K}, a^{\dagger}. \rbrack &=& (-i \omega - \gamma / 2)\hskip 0.1cm  a^{\dagger}. + \gamma \hskip 0.1cm .a^{\dagger} \, , \nonumber \\ 
\lbrack{\cal K}, .a^{\dagger} \rbrack &=&   (-i \omega +  \gamma / 2) \hskip 0.1cm .a^{\dagger} \, , \nonumber \\
\lbrack {\cal K}, .a \rbrack &=&  (i \omega - \gamma / 2 )\hskip 0.1cm  .a + \gamma \hskip 0.1cm a. \, , \nonumber \\ 
\lbrack {\cal K}, a. \rbrack &=&   (i \omega + \gamma / 2) \hskip 0.1cm a. \hskip 0.3cm .
\end{eqnarray}
We may obtain from linear combinations of these relations, new superoperators, 

\begin{eqnarray}\label{superoperadores}
{\cal M}_+ = a^{\dagger} . - .a^{\dagger} \hskip 0.2cm &,&  \hskip 0.2cm {\cal M}_- =  a. \, ,  \nonumber \\ 
{\cal N}_+ = .a - a.  \hskip 0.2cm &,&  \hskip 0.2cm {\cal N}_- =  .a^{\dagger} \, , 
\end{eqnarray}
which satisfy simpler commutation relations with ${\cal K},$ 

\begin{eqnarray}\label{comutadorK}
\lbrack{\cal K},{\cal M}_{\pm}] &=& \pm (-i \omega - \gamma /2){\cal M}_{\pm} \, , \nonumber \\
\lbrack {\cal K},{\cal N}_{\pm}\rbrack &=& \pm (i \omega - \gamma /2){\cal N}_{\pm} \, .
\end{eqnarray}
Therefore ${\cal M}_{\pm}$ and ${\cal N}_{\pm}$ are ladder superoperators. The definition of these ladder superoperators was possible because the commutation relations (\ref{comutadora}) correspond to a closed algebra. In fact, this technique is limited to cases in which the commutation relations between the Liouvillian and the superoperators that constitute it form a closed algebra. 

The ladder superoperators were normalized so that they satisfy commutation relations identical to those of two decoupled harmonic oscillators,

\begin{eqnarray}\label{comutadorsuper}
\lbrack{\cal M}_-, {\cal M}_+] &=& 1  \hskip 0.2cm ,  \hskip 0.2cm \lbrack {\cal N}_-, {\cal N}_+ \rbrack  = 1 \, , \nonumber \\
\lbrack {\cal M}_{+}, {\cal N}_{+} \rbrack &=& 0 \hskip 0.2cm ,  \hskip 0.2cm \lbrack {\cal M}_{+}, {\cal N}_{-}\rbrack = 0 \, , \nonumber \\
\lbrack {\cal M}_{-}, {\cal N}_{+}\rbrack  &=& 0 \hskip 0.2cm ,  \hskip 0.2cm \lbrack {\cal M}_{-}, {\cal N}_{-}\rbrack = 0 \, .
\end{eqnarray}

Using these superoperators we may rewrite the Liouvillian ${\cal K}$ in the suggestive form,

\begin{eqnarray}\label{K}
{\cal K} = (-i \omega - \gamma/2) {\cal M}_+ {\cal M}_- + (i \omega - \gamma/2) {\cal N}_+ {\cal N}_- \, .
\end{eqnarray}

Due to the fact that ${\cal M}_+ {\cal M}_-$ and ${\cal N}_+ {\cal N}_-$ commute, we can find simultaneous eigenstates of these superoperators, i.e.,

\begin{eqnarray}\label{autovalor}
{\cal M}_+{\cal M}_- \hskip 0.1cm R^{m,n} &=& m \hskip 0.1cm R^{m,n}\, , \nonumber \\
{\cal N}_+ {\cal N}_- \hskip 0.1cm R^{m,n} &=& n \hskip 0.1cm R^{m,n}\, .
\end{eqnarray}
It is clear that these eigenstates will be also eigenstates of the Liouvillian,

\begin{equation}\label{autovalorK}
{\cal K} \hskip 0.1cm R^{m,n} = [ m(-i \omega - \gamma/2) + n(i \omega - \gamma/2)] R^{m,n}\, .
\end{equation}

Since the density operator determines the ensemble average of observables by means of traces, $\langle A(t) \rangle = {\rm tr} (A \rho(t)),$ we must assume that it has finite trace with any other operator. More generally, the density operator should have finite trace with any superoperator. Since the eigenstates $R^{m,n}$ will be used in the expansion of the density operator, we must impose that they also have this property. Therefore, solutions of Eq.~(\ref{autovalor}) like $1$ (with eigenvalues $m=-1$ and $n=-1),$ $a$ $(m=-2,n=-1),$ $a^{\dagger}$ $(m=-1, n=-2)$ and $aa^{\dagger}$ $(m=-2, n=-2)$ cannot represent $R^{m,n}$'s, because they have divergent trace with, for example, the superoperators $1.,$ $a^{\dagger}.,$ $a.$ and $1.,$ respectively. This shows that negative eigenvalues do generate unphysical solutions. In fact, these solutions correspond to density operators that grow without bound as $t$ increases, which is not physically expected in the case of a zero-temperature reservoir. Therefore, we will only attempt solutions of Eq.~(\ref{autovalor}) with positive eigenvalues.

From the commutation relations (\ref{comutadorsuper}) and from Eq.~(\ref{autovalor}) we can show that ${\cal M}_+$ increases the eigenvalue of ${\cal M}_+ {\cal M}_-$ by one unity and does not change the eigenvalue of ${\cal N}_+ {\cal N}_-.$ In analogous form we have that ${\cal N}_+$ only changes the eigenvalue of ${\cal N}_+ {\cal N}_-.$ So far we have not specify any type of normalization for the eigenstates $R^{m,n}.$ Therefore, we may choose the coefficients of the step-up relations as being $\sqrt{m+1}$ and $\sqrt{n+1},$ in analogy with the harmonic oscillator, so that the step-up relations have the form 

\begin{eqnarray}\label{levantamento}
{\cal M}_+ \hskip 0.1cm R^{m,n} &=& \sqrt{m+1} \hskip 0.1cm R^{m+1,n}\, , \nonumber \\
{\cal N}_+ \hskip 0.1cm R^{m,n} &=& \sqrt{n+1} \hskip 0.1cm R^{m,n+1} \, .
\end{eqnarray}
From relations (\ref{comutadorsuper}), (\ref{autovalor}) and (\ref{levantamento}) we obtain the step-down relations

\begin{eqnarray}\label{abaixamento}
{\cal M}_- \hskip 0.1cm R^{m,n} &=& \sqrt{m} \hskip 0.1cm R^{m-1,n}\, , \nonumber \\
{\cal N}_- \hskip 0.1cm R^{m,n} &=& \sqrt{n} \hskip 0.1cm R^{m,n-1} \, .
\end{eqnarray}

We will look for the eigenstate of null eigenvalues, i.e., $R^{0,0}.$ The above relations show that we cannot obtain new eigenstates by applying ${\cal M}_-$ and ${\cal N}_-$ to $R^{0,0},$     

\begin{eqnarray}\label{abaixa0}
{\cal M}_- \hskip 0.1cm R^{0,0} = a R^{0,0} = 0 \hskip 0.2cm ,  \hskip 0.2cm {\cal N}_- \hskip 0.1cm R^{0,0} = R^{0,0}a^{\dagger} = 0 \, .
\end{eqnarray}

Although the convention used in the step-up relations gives a restriction in the normalization of the eigenstates, we still have the freedom to choose the normalization of one of them, say $R^{0,0}.$ We will choose it such that 

\begin{equation}\label{T1}
{\rm tr} R^{0,0} = 1 \, .
\end{equation}
Thus, this eigenstate takes the simple form

\begin{equation}\label{R00}
R^{0,0} =  \vert 0 \rangle \langle 0 \vert \, ,
\end{equation}
where $\vert 0 \rangle$ represents the oscillator ground state. The remaining eigenstates with positive integer eigenvalues, $m,n = 0,1,2,\dots,$ may be then obtained by applying ${\cal M}_+$ and ${\cal N}_+$ successively to $R^{0,0},$

\begin{equation}\label{Rmn}
R^{m,n} = \frac{{\cal N}_+^n }{ \sqrt{n!}} \frac{{\cal M}_+^m }{ \sqrt{m!}} R^{0,0}\, .
\end{equation} 
Thus, we may obtain the explicit form of the eigenstates in the Fock basis

\begin{eqnarray}\label{autoestado}
R^{m,n} = \sum_{k = 0}^{{\rm min}(m,n)} \frac{(-1)^k }{ k!}  \sqrt{\frac{m! }{  (m - k)!} \frac{n! }{  (n - k)!}} \nonumber \\  
\times \vert m - k \rangle \langle n - k\vert \, .
\end{eqnarray}

Using the cyclic property of the trace, we get ${\rm tr}({\cal M}_+ \hskip 0.1cm \rho)=0$ and ${\rm tr}({\cal N}_+ \hskip 0.1cm \rho)=0$ for any target operator $\rho.$ Thus, taking the trace of Eq.~(\ref{autovalor}), we obtain that $m \hskip 0.1cm {\rm tr} R^{m,n} = 0$ and  $n \hskip 0.1cm {\rm tr} R^{m,n} = 0.$ This allows us to conclude that the trace of $R^{m,n}$ will be identically null for $m$ or $n$ different from zero. Taking into account the convention (\ref{T1}), we may write

\begin{equation}\label{TRmn}
{\rm tr} R^{m,n} = \delta_{m,0}  \hskip 0.1cm  \delta_{n,0} \, .
\end{equation}
By decreasing the eigenvalues of a given eigenstate,

\begin{equation}\label{abaixaRmn}
  \frac{{\cal N}_-^j }{ \sqrt{j!}}  \frac{{\cal M}_-^i }{ \sqrt{i!}} \hskip 0.1cm R^{m,n} = \sqrt{\frac{m!}{i!(m-i)!} \frac{n!}{j!(n-j)!}}R^{m-i,n-j}\, ,
\end{equation}
and taking the trace of this equation, we obtain a very useful property,

\begin{equation}\label{ortogonal}
{\rm tr} \biggl( \frac{{\cal N}_-^j }{ \sqrt{j!}} \frac{ {\cal M}_-^i }{ \sqrt{i!}} \hskip 0.1cm R^{m,n} \biggr) = \delta_{m,i} \delta_{n,j}\, .
\end{equation}

It is shown in the appendix that the $R^{m,n}$'s with positive integer eigenvalues, $m,n = 0,1,2,\dots,$ form a complete set, which allows us to expand any density operator into this eigenstates. In particular we may expand $\rho(0),$

\begin{equation}\label{expansao}
\rho (0) = \sum_{m=0}^{\infty}\sum_{n=0}^{\infty}  C^{m,n} \hskip 0.1cm R^{m,n} \, .
\end{equation}
Using Eq.~(\ref{ortogonal}), we easily obtain the expansion coefficients,

\begin{equation}\label{Cmn}
C^{m,n} = {\rm tr} \biggl(\frac{{\cal N}_-^n }{ \sqrt{n!}}  \frac{{\cal M}_-^m }{ \sqrt{m!}}\hskip 0.1cm \rho (0)  \biggr)\, , 
\end{equation}
or more explicitly,

\begin{equation}\label{Cmnexplicito}
C^{m,n} = \frac{1 }{ \sqrt{m!n!}} \hskip 0.1cm {\rm tr} (a^m  \rho (0) a^{\dagger \hskip 0.1cm n})\, .
\end{equation}

The solution of the master equation will be obtained by substituting Eq.~(\ref{expansao}) into Eq.~(\ref{formal}) and using Eq.~(\ref{autovalorK}), 

\begin{eqnarray}\label{solucao}
\rho (t) &=& \sum_{m=0}^{\infty}\sum_{n=0}^{\infty} C^{m,n} \hskip 0.1cm R^{m,n} \nonumber \\ &\times& e^{\displaystyle{[m(-i\omega - \gamma / 2) + n (i \omega - \gamma / 2)] t}}\, .
\end{eqnarray}
Thus, to obtain the solution $\rho(t)$ of a given initial condition, $\rho(0),$ we should determine the coefficients $C^{m,n}$ and perform the summations over the indices $m$ and $n$ in Eq.~(\ref{solucao}), using the explicit form of the eigenstates $R^{m,n}.$ 

Notice that $C^{0,0} = {\rm tr} \rho(0) = 1,$ independently of the initial condition. Thus, from expansion (\ref{solucao}), it is clear that any initial state decays to the eigenstate $R^{0,0}$ as $t$ goes to infinity. This is due to the fact that the system dissipates energy to the reservoir until the system reaches its ground state. 

As an example, we will use as initial condition the oscillator in a Fock state,

\begin{equation}\label{R0N}
\rho_{N}(0) = \vert N \rangle \langle N \vert\, ,
\end{equation}
where the expansion coefficients are given by

\begin{equation}\label{CmnN}
C^{m,n}_N = \delta_{m,n} \frac{N! }{ n! (N - n)!}\, . 
\end{equation}
By performing the summations in Eq.~(\ref{solucao}), we will obtain the density operator at time $t,$

\begin{equation}\label{RtN}
\rho(t)_N = \sum_{k=0}^{N} \frac{N! }{ k! (N - k)!} e^{\displaystyle{- k \gamma t}} (1 - e^{\displaystyle{- \gamma t}})^{N - k} \vert k \rangle \langle k \vert \, .
\end{equation}

We can also easily calculate the density operator when the oscillator is initially in a coherent state,

\begin{equation}\label{R0a}
\rho_{\alpha}(0) = \vert \alpha \rangle \langle \alpha \vert \, .
\end{equation}
In this case, we have for the expansion coefficients

\begin{equation}\label{Cmnalfa}
C^{m,n}_{\alpha} = \frac{\alpha^m }{ \sqrt{m!}} \frac{\alpha^{*n} }{ \sqrt{n!}}
\end{equation}
and then the density operator will be given by

\begin{equation}\label{Rta}
\rho_{\alpha}(t) = \vert \alpha e^{(-i \omega - \gamma/2)t} \rangle \langle \alpha e^{(-i \omega - \gamma/2)t} \vert \, ,  
\end{equation}
indicating that the oscillator dissipates but remains as a pure state during all the evolution.

\section{Harmonic oscillator in a thermal bath}

We now consider the case of a harmonic oscillator in contact with a thermal reservoir. The Hamiltonian of the system is the same as that given by (\ref{Ha}), but the Lindblad superoperator is now given by \cite{milburn,miguel}

\begin{eqnarray}\label{L}
{\cal L} &=& (\bar n + 1)\frac{\gamma }{ 2} (2 a..a^{\dagger} - a^{\dagger} a. - .a^{\dagger} a) \nonumber \\
&+& \bar n \frac{\gamma }{ 2} (2 a^{\dagger}..a - a a^{\dagger}. -.a a^{\dagger})\, ,
\end{eqnarray}
where $\bar n = ( e^{\hskip 0.1cm \beta \displaystyle{\hbar \omega }} - 1)^{-1}$ is the average number of quanta in the reservoir, $\beta = 1 / k_BT,$ being $k_B$ the Boltzmann's constant. For this case, we will define new step-up and step-down superoperators,

\begin{eqnarray}\label{superoperadoresT}
\widetilde{\cal M}_+ = a^{\dagger} . - .a^{\dagger} \hskip 0.2cm &,&  \hskip 0.2cm \widetilde{\cal M}_- = - \hskip 0.1cm \bar n \hskip 0.1cm .a +(\bar n + 1) \hskip 0.1cm a. \, , \nonumber \\
\widetilde{\cal N}_+ = .a - a. \hskip 0.2cm &,&  \hskip 0.2cm \widetilde{\cal N}_- = - \hskip 0.1cm \bar n \hskip 0.1cm a^{\dagger}. + (\bar n +1 )\hskip 0.1cm .a^{\dagger} \hskip 0.3cm .
\end{eqnarray}
Using these superoperators the new Liouvillian $\widetilde{\cal K}$ may be rewritten in a form analogous to Eq.(\ref{K}),

\begin{eqnarray}\label{KT}
\widetilde{\cal K} = (-i \omega - \gamma/2) \widetilde{\cal M}_+ \widetilde{\cal M}_- + (i \omega - \gamma/2) \widetilde{\cal N}_+ \widetilde{\cal N}_- \, .
\end{eqnarray}

As the superoperators $\widetilde{\cal M}_+,$ $\widetilde{\cal M}_-,$ $\widetilde{\cal N}_+,$ and $\widetilde{\cal N}_-$ obey the same commutation rules as those obeyed by the superoperators ${\cal M}_+,$ ${\cal M}_-,$ ${\cal N}_+,$ and ${\cal N}_-,$ we follow the same procedure used in the last section to obtain the simultaneous eigenstates of  $\widetilde{\cal M}_+ \widetilde{\cal M}_-$ and $\widetilde{\cal N}_+ \widetilde{\cal N}_-.$ We denote these eigenstates by $R^{m,n}(\bar n).$ The eigenstate $R^{0,0}(\bar n)$ must have unit trace and obey the equations $\widetilde{\cal M}_- \hskip 0.1cm R^{0,0}(\bar n) = 0$ and $\widetilde{\cal N}_- \hskip 0.1cm R^{0,0}(\bar n) = 0,$ which will give us

\begin{equation}\label{R00T}
R^{0,0}(\bar n) = \sum_{k=0}^{\infty} \hskip 0.1cm \frac{ \bar n^k }{ \hskip 0.1cm (\bar n + 1)^{k + 1}} \hskip 0.1cm \vert k \rangle \langle k \vert \, .
\end{equation}
This operator exactly represents a thermal distribution, $R^{0,0}(\bar n) = e^{- \beta \displaystyle{H}}/ Z,$ being $Z = {\rm tr} e^{- \beta \displaystyle{H}}$ the partition function of the problem, since this state is in thermal equilibrium with the reservoir. The remaining eigenstates with positive integer eigenvalues, $m,n = 0,1,2,\dots,$ may be obtained by applying $\widetilde{\cal M}_+$ and $\widetilde{\cal N}_+$ successively to $R^{0,0}(\bar n),$ 

\begin{equation}
R^{m,n}(\bar n) = \frac{\widetilde{\cal N}_+^n }{ \sqrt{n!}} \frac{\widetilde{\cal M}_+^m }{ \sqrt{m!}} R^{0,0}(\bar n)\, ,
\end{equation}
which results in

\begin{eqnarray}\label{autoestadoT}
& &R^{m,n}(\bar n)  \nonumber \\
&=& \cases{\displaystyle{\sum_{k = 0}^{\infty}} \hskip 0.1cm \sqrt{\frac{n! k! }{ m! (k + m -n)!}} \hskip 0.1cm \frac{1 }{ (\bar n + 1)^{m+1}} \cr \cr 
\times P_{\hskip 0.1cm \hskip 0.1cm m}^{\hskip 0.1cm k,k -n} \biggl(\displaystyle{\frac{\bar n }{ \bar n +1}}\biggr) \vert k + m -n \rangle \langle k\vert \hskip 0.1cm , \hskip 0.1cm m \geq n \cr  \cr \cr
\displaystyle{\sum_{k = 0}^{\infty}} \hskip 0.1cm \sqrt{\frac{m! k! }{ n! (k + n -m)!}} \hskip 0.1cm \frac{1 }{ (\bar n + 1)^{n+1}} \cr \cr 
\times P_{\hskip 0.1cm \hskip 0.1cm n}^{\hskip 0.1cm k,k-m} \biggl(\displaystyle{\frac{\bar n }{ \bar n +1}}\biggr) \vert k\rangle  \langle k +n - m \vert \hskip 0.1cm , \hskip 0.1cm m \leq n \, ,\cr}
\end{eqnarray}
where $P_{\hskip 0.1cm \hskip 0.1cm m}^{\hskip 0.1cm k,l}(x)$ are the polynomials

\begin{equation}
P_{\hskip 0.1cm \hskip 0.1cm m}^{\hskip 0.1cm k,l}(x) = \sum_{j={\rm max} (0,l)}^{k} (-1)^{j-l} \frac{(j + m)! }{ (j-l)! (k-j)!} \frac{x^j }{ j!}\, .
\end{equation}

Following the procedure adopted in the last section, the solution of the master equation for the case of a harmonic oscillator in a thermal bath takes the form

\begin{eqnarray}\label{solucaoT}
\rho_T (t) &=& \sum_{m=0}^{\infty}\sum_{n=0}^{\infty} C^{m,n}(\bar n) \hskip 0.1cm R^{m,n}(\bar n) \nonumber \\
&\times& e^{\displaystyle{[m(-i\omega - \gamma / 2) + n (i \omega - \gamma / 2)] t}}\, ,
\end{eqnarray}
where the expansion coefficients are obtained in a form analogous to Eq.(\ref{Cmn}),

\begin{equation}\label{CmnT}
C^{m,n}(\bar n) =  {\rm tr} \biggl(\frac{\widetilde{\cal N}_-^{\hskip 0.1cm n} }{ \sqrt{n!}}  \frac{\widetilde{\cal M}_-^{\hskip 0.1cm m} }{ \sqrt{m!}}\hskip 0.1cm \rho (0)  \biggr)\, . 
\end{equation}

As a simple example, we consider an oscillator which is initially in a thermal state at temperature $T_0$ different from the reservoir temperature $T$,

\begin{equation}\label{R0T}
\rho_{T}(0) = \sum_{k=0}^{\infty} \frac{\bar n_0^k }{ (\bar n_0 + 1)^{k+1}}\vert k \rangle \langle k \vert \, ,
\end{equation}
where $\bar n_0 = (e^{\beta_0 \displaystyle{\hbar \omega}} - 1)^{-1},$ being $\beta_0 = 1/ k_B T_0.$ In this case, the expansion coefficients take the very simple form,

\begin{equation}\label{CmnT}
C^{m,n} = \delta_{m,n} \hskip 0.1cm (\bar n_0 - \bar n)^n \, .
\end{equation}
The solution of the master equation is obtained substituting the above result into (\ref{solucaoT}), rearranging the summations and using the identity \cite{identity}

\begin{eqnarray}\label{identidade}
\sum_{n=k}^{\infty} (-1)^{n} \frac{n! }{ (n-k)!} x^{n-k} = (-1)^k \frac{k! }{ (1+x)^{k+1}}  \, .
\end{eqnarray}
So that the density operator remains as a thermal distribution during all the time evolution, 

\begin{equation}\label{RtT}
\rho_T(t) = \sum_{k=0}^{\infty} \hskip 0.1cm \frac{ \bar n(t)^k }{ \hskip 0.1cm (\bar n(t) + 1)^{k + 1}} \hskip 0.1cm \vert k \rangle \langle k \vert \, ,   
\end{equation}
where $\bar n(t) = \bar n + (\bar n_0 - \bar n) e^{\displaystyle{- \gamma t}},$ showing that the oscillator reaches the thermal equilibrium with the reservoir in a characteristic time $1 / \gamma.$

\section{Two-level system in a thermal bath}

Another interesting example is a two-level system, with a ground state $\vert g \rangle$ and a excited state $\vert e\rangle,$ in contact with a thermal reservoir. In this case, we must solve a master equation of type (\ref{eqm}), where the Hamiltonian of the system is given by

\begin{equation}\label{Hfermionico}
H = \hbar \omega \sigma_z/2 \, ,
\end{equation}
being $\sigma_z = \vert e\rangle \langle e \vert - \vert g\rangle \langle g \vert,$ and the Lindblad superoperator is given by \cite{milburn,miguel} 

\begin{eqnarray}\label{Lfermionico}
{\cal L} &=& (\bar n + 1)\frac{\gamma }{ 2} (2 \sigma_- .. \sigma_+ - \sigma_+ \sigma_- . - . \sigma_+ \sigma_-) \nonumber \\
&+& \bar n \frac{\gamma }{ 2} (2 \sigma_+ .. \sigma_- - \sigma_- \sigma_+. - . \sigma_- \sigma_+)\, ,
\end{eqnarray}
where $\sigma_+ = \vert e \rangle \langle g \vert,$ $\sigma_- = \vert g \rangle \langle e \vert$ and $\bar n = (e^{\beta \displaystyle{\hbar \omega}} - 1)^{-1}.$ The master equation for this problem can be easily solved in the basis of the states $\vert e \rangle$ and $\vert g \rangle.$ Although we are aware of this, we proceed to solve this problem with our method to give a simple example of dealing with anticommutation relations. 

The Lindblad superoperator can be rewritten in a more convenient form,

\begin{eqnarray}\label{LN}
{\cal L} &=& (1 - \bar N)\frac{\Gamma }{ 2} (2 \sigma_- .. \sigma_+ - \sigma_+ \sigma_- . - . \sigma_+ \sigma_-) \nonumber \\
&+& \bar N \frac{\Gamma }{ 2} (2 \sigma_+ .. \sigma_- - \sigma_- \sigma_+. - . \sigma_- \sigma_+)\, ,
\end{eqnarray}
where $\bar N = (e^{\beta \displaystyle{\hbar \omega}} + 1)^{-1}$ and $\Gamma = \gamma \hskip 0.1cm \coth (\beta \hbar \omega /2).$ For this case, we will define the step-up and step-down superoperators

\begin{eqnarray}\label{superoperadoresfermionico}
{\cal P}_+ &=& \sigma_+ . + \sigma_z .. \sigma_+ \, , \nonumber \\
{\cal P}_- &=& (1 - \bar N) \sigma_- . + \bar N \sigma_z .. \sigma_- \, , \nonumber \\
{\cal Q}_+ &=& .\sigma_-  + \sigma_- .. \sigma_z \, , \nonumber \\
{\cal Q}_-  &=& (1- \bar N) .\sigma_+  + \bar N \sigma_+ .. \sigma_z  \hskip 0.3cm .
\end{eqnarray}

By defining the anticommutator between superoperators in the usual form, $\{{\cal A},{\cal B}\} := {\cal A}{\cal B} + {\cal B}{\cal A},$ we will have the following anticommutation relations:

\begin{eqnarray}\label{anticomutadores}
\{{\cal P}_-,{\cal P}_+ \} = {\rm 1\!\!\hskip 1 pt l} \hskip 0.2cm &,&  \hskip 0.2cm \{{\cal Q}_-,{\cal Q}_+ \} = {\rm 1\!\!\hskip 1 pt l} \, , \nonumber \\  
\{{\cal P}_{+},{\cal P}_{+} \} = 0 \hskip 0.2cm &,&  \hskip 0.2cm \{{\cal Q}_{+},{\cal Q}_{+} \} = 0 \, , \nonumber \\
\{{\cal P}_{-},{\cal P}_{-} \} = 0 \hskip 0.2cm &,&  \hskip 0.2cm \{{\cal Q}_{-},{\cal Q}_{-} \} = 0 \, ,
\end{eqnarray}
and the following commutation relations:

\begin{eqnarray}\label{comutacaospin}
[{\cal P}_{+},{\cal Q}_{+} \rbrack = 0 \hskip 0.2cm &,&  \hskip 0.2cm \lbrack {\cal P}_{+},{\cal Q}_{-} \rbrack = 0 \, , \nonumber \\
\lbrack {\cal P}_{-},{\cal Q}_{+} \rbrack = 0 \hskip 0.2cm &,&  \hskip 0.2cm \lbrack {\cal P}_{-},{\cal Q}_{-} \rbrack = 0 \, , 
\end{eqnarray}
where ${\rm 1\!\!\hskip 1 pt l} = \vert e \rangle \langle e \vert + \vert g \rangle \langle g \vert.$ We may again rewrite ${\cal K}$ in terms of these superoperators, 

\begin{eqnarray}\label{Kfermionico}
{\cal K} = (-i \omega - \Gamma/2) {\cal P}_+ {\cal P}_- + (i \omega - \Gamma/2) {\cal Q}_+ {\cal Q}_- \, ,
\end{eqnarray}
and find simultaneous eigenstates of ${\cal P}_+{\cal P}_-$ and ${\cal Q}_+{\cal Q}_-,$ due to the fact that they commute,

\begin{eqnarray}\label{autovalorfermionico}
{\cal P}_+{\cal P}_- \hskip 0.1cm S^{p,q} &=& p \hskip 0.1cm S^{p,q}\, , \nonumber \\
{\cal Q}_+ {\cal Q}_- \hskip 0.1cm S^{p,q} &=& q \hskip 0.1cm S^{p,q} \, .
\end{eqnarray}

The anticommutation relations enable us to obtain the commutation relation $[ {\cal P}_+ {\cal P}_- , {\cal P}_- ] = - {\cal P}_- .$ Applying this relation on the eigenstate $S^{p,q}$ and using ${\cal P}_-^2 = 0,$ we will have $(p-1){\cal P}_- S^{p,q} = 0.$ Then, acting ${\cal P}_+$ on this last expression, we will obtain $p(p-1)S^{p,q} = 0,$ allowing us to conclude that, if we want $S^{p,q} \neq 0,$ we must have $p=0,1.$ We may proceed in analogous form and obtain $q=0,1.$

Here we also have step-up and step-down relations, which we will not explicit. The eigenstate $S^{0,0}$ is obtained from equations ${\cal P}_- S^{0,0} = 0,$ ${\cal Q}_- S^{0,0} = 0$ and ${\rm tr} S^{0,0} = 1.$ Then we find

\begin{equation}
S^{0,0} = \bar N \vert e \rangle \langle e \vert + (1 - \bar N)\vert g \rangle \langle g \vert \, ,
\end{equation}
which also can be identified as a thermal distribution $S^{0,0} = e^{- \beta \displaystyle{ H}}/Z,$ where $H= \hbar \omega \sigma_z/2$ and $Z = {\rm tr} e^{-\beta \displaystyle{H}}$ is the partition function of the problem. The remaining eigenstates will be obtained by stepping $S^{0,0}$ up, 

\begin{equation}
S^{p,q} = {\cal Q}^q_{+} {\cal P}^p_{+} S^{0,0} \hskip 0.2cm , \hskip 0.2cm p,q = 0,1 \, ,
\end{equation}
which will give us

\begin{eqnarray}
S^{1,0} &=& \vert e \rangle \langle g \vert \hskip 0.2cm , \hskip 0.2cm S^{0,1} = \vert g \rangle \langle e \vert \, , \nonumber \\  
S^{1,1} &=& \vert e \rangle \langle e \vert  - \vert g \rangle \langle g \vert \, .
\end{eqnarray}

Obviously, the basis formed by four eigenstates $S^{p,q}$ is complete since, in this case, the space of operators is four-dimensional.

Following the procedure adopted in the first section, the solution of the new master equation takes the form

\begin{equation}\label{expansaofermionico}
\rho (t) = \sum_{p=0}^{1}\sum_{q=0}^{1}  C^{p,q} \hskip 0.1cm S^{p,q} e^{\displaystyle{[p(-i\omega - \Gamma/2) + q(i\omega - \Gamma/2)]t}}\, ,
\end{equation}
where the expansion coefficients are obtained from expression

\begin{equation}\label{Cmnfermionico}
C^{p,q} = {\rm tr} ({\cal Q}_-^q   {\cal P}_-^p \hskip 0.1cm \rho (0)) \, .
\end{equation}

It is interesting to note that there is more than one choice for ladder superoperators. For example, the superoperators

\begin{eqnarray}\label{superoperadorspin}
{\cal P}_+^{\prime} &=& .\sigma_+  - \sigma_+ .. \sigma_z \, , \nonumber \\
{\cal P}_-^{\prime} &=& \bar N .\sigma_-  - (1 - \bar N) \sigma_- .. \sigma_z \, , \nonumber \\
 {\cal Q}_+^{\prime} &=& \sigma_-.  - \sigma_z .. \sigma_- \, , \nonumber \\
{\cal Q}_-^{\prime} &=& \bar N \sigma_+.  - (1- \bar N) \sigma_z .. \sigma_+ \, ,
\end{eqnarray}
obey anticommutation relations identical to (\ref{anticomutadores}) and commutation relations identical to (\ref{comutacaospin}). Furthermore, these superoperators rewrite the Liouvillian in a form which is analogous to (\ref{Kfermionico}), due to equalities ${\cal P}_+^{\prime} {\cal P}_-^{\prime} = {\cal P}_+ {\cal P}_-$ and ${\cal Q}_+^{\prime} {\cal Q}_-^{\prime} = {\cal Q}_+ {\cal Q}_-.$ However, the superoperators ${\cal P}_+,$ ${\cal P}_-,$ ${\cal Q}_+^{\prime}$ and ${\cal Q}_-^{\prime}$ have anticommutation relations identical to (\ref{anticomutadores}) but, instead of obeying the commutation relations given in (\ref{comutacaospin}), they obey anticommutation relations. An analogous fact occurs with the group of superoperators ${\cal P}_+^{\prime},$ ${\cal P}_-^{\prime},$ ${\cal Q}_+$ and ${\cal Q}_-.$ 

As a simple example, we can use again as initial condition the system in a thermal state at temperature $T_0$ different from the reservoir temperature,

\begin{equation}
\rho(0) =  \bar N_0 \vert e \rangle \langle e \vert + (1 - \bar N_0)\vert g \rangle \langle g \vert \, ,
\end{equation}
where $\bar N_0 = (e^{\beta_0 \displaystyle{\hbar \omega}} + 1)^{-1}.$ In this case, we will have for expansion coefficients

\begin{equation}
C^{p,q}= \delta_{p,q}(\bar N_0 - \bar N)^q \, ,
\end{equation}
so that the density operator at time $t$ will be given by 

\begin{equation}
\rho(t) = \bar N(t) \vert e \rangle \langle e \vert + (1 - \bar N(t))\vert g \rangle \langle g \vert \, ,
\end{equation}
where $\bar N(t) = \bar N + (\bar N_0 - \bar N)e^{\displaystyle{-\Gamma t}},$ i.e., the system remains as a thermal distribution during all the evolution, reaching the thermal equilibrium with the reservoir in the characteristic time $1/\Gamma.$

\section{Two coupled modes in contact with a zero-temperature reservoir}

We will analyze now the case of two coupled harmonic oscillators in contact with a zero-temperature reservoir. In this case, the master equation is still of type (\ref{eqm}), but now the Hamiltonian of the system is given by

\begin{eqnarray}\label{Hab}
H = \hbar \omega_a a^{\dagger} a + \hbar \omega_b b^{\dagger} b + \hbar g a^{\dagger} b + \hbar g^* a b^{\dagger}
\end{eqnarray}
and the Lindblad of this problem is 

\begin{eqnarray}\label{Ldoismodos}
{\cal L} &=& \frac{\gamma_a}{2} (2 a..a^{\dagger} - a^{\dagger} a. - .a^{\dagger} a) \nonumber \\
&+& \frac{\gamma_b }{ 2} (2 b..b^{\dagger} - b^{\dagger} b. - .b^{\dagger} b) \nonumber \\
&+& \frac{\gamma_c }{ 2} (2 b..a^{\dagger} - a^{\dagger} b. - .a^{\dagger} b) \nonumber \\
&+&  \frac{\gamma_c^* }{ 2} (2 a..b^{\dagger} - b^{\dagger} a. - .b^{\dagger} a)\, ,
\end{eqnarray}
where $\vert \gamma_c  \vert = \sqrt{\gamma_a \gamma_b}$ for the case in which the reservoir is common to the two modes and $\gamma_c =0$ for the case in which there are separate reservoirs for each mode \cite{sonia}. For both cases we will define a set of step-up superoperators

\begin{eqnarray}\label{levantamentoab}
{\cal M}_+ &=& r_+ {\cal M}_+^a + s_+ {\cal M}_+^b \, , \nonumber \\
{\cal N}_+ &=& r_- {\cal M}_+^a + s_- {\cal M}_+^b \, , \nonumber \\
{\cal P}_+ &=& r_+^* {\cal N}_+^a + s_+^* {\cal N}_+^b \, , \nonumber \\
{\cal Q}_+ &=& r_-^* {\cal N}_+^a + s_-^* {\cal N}_+^b \, , \nonumber \\
\end{eqnarray}
and step-down superoperators

\begin{eqnarray}\label{abaixamentoab}
{\cal M}_- &=& u_+ {\cal M}_-^a + v_+ {\cal M}_-^b \, , \nonumber \\
{\cal N}_- &=& u_- {\cal M}_-^a + v_- {\cal M}_-^b \, , \nonumber \\
{\cal P}_- &=& u_+^* {\cal N}_-^a + v_+^* {\cal N}_-^b \, , \nonumber \\
{\cal Q}_- &=& u_-^* {\cal N}_-^a + v_-^* {\cal N}_-^b \, ,
\end{eqnarray}
where the superoperators with indices $a,b$ refer to those of a single mode of the form (\ref{superoperadores}) (for example, ${\cal M}_-^a = a.$ and ${\cal M}_-^b =  b.).$ The coefficients in (\ref{levantamentoab}) and (\ref{abaixamentoab}) are defined as

\begin{eqnarray}\label{coeficientes}
r_{\pm} &=& i \frac{S \pm \Delta }{ \sqrt{2i \Delta (S \pm \Delta)}} \hskip 0.2cm , \hskip 0.2cm s_{\pm} = i \frac{V }{ \sqrt{2i \Delta (S \pm \Delta)}}\, , \nonumber \\
u_{\pm} &=& \pm \frac{S \pm \Delta }{ \sqrt{2i \Delta (S \pm \Delta)}} \hskip 0.2cm , \hskip 0.2cm v_{\pm} = \pm \frac{U }{ \sqrt{2i \Delta (S \pm \Delta)}} \, ,
\end{eqnarray}
where the parameters $U,$ $V,$ $S$ and $\Delta$ are given by

\begin{eqnarray}\label{parametros}
U &=& g  - i \frac{\gamma_c }{ 2} \hskip 0.2cm , \hskip 0.2cm V = g^* - i \frac{\gamma_c^* }{ 2} \, , \nonumber \\
S &=& \frac{\omega_a - \omega_b }{ 2} - i \frac{\gamma_a - \gamma_b }{ 4} \, , \nonumber \\
\Delta &=& \sqrt{S^2 + UV} \, .
\end{eqnarray}
These superoperators obey the commutation relations 

\begin{eqnarray}\label{comutacaoab}
\lbrack{\cal M}_-, {\cal M}_+\rbrack = 1 \hskip 0.2cm &,&  \hskip 0.2cm \lbrack{\cal N}_-, {\cal N}_+\rbrack = 1 \, , \nonumber \\
\lbrack{\cal P}_-, {\cal P}_+\rbrack = 1 \hskip 0.2cm &,&  \hskip 0.2cm \lbrack{\cal Q}_-, {\cal Q}_+\rbrack = 1 \, , 
\end{eqnarray}
whereas the remaining relations are identically null. The Liouvillian ${\cal K}$ can be rewritten in terms of these superoperators as

\begin{eqnarray}\label{Kab}
{\cal K} &=& \lambda_+ {\cal M}_+ {\cal M}_- + \lambda_- {\cal N}_+ {\cal N}_- \nonumber \\
&+& \lambda_+^* {\cal P}_+ {\cal P}_- + \lambda_-^* {\cal Q}_+ {\cal Q}_- \, ,
\end{eqnarray}
where

\begin{eqnarray}\label{lambda}
\lambda_{\pm} = - i R \mp i \Delta  \hskip 0.2cm , \hskip 0.2cm R = \frac{\omega_a + \omega_b }{ 2} - i \frac{\gamma_a + \gamma_b }{ 4}\, .
\end{eqnarray}

Following the procedure used before, we may find the simultaneous eigenstates of ${\cal M}_+ {\cal M}_-,$ ${\cal N}_+ {\cal N}_-,$ ${\cal P}_+ {\cal P}_-$ and ${\cal Q}_+ {\cal Q}_-$ with eigenvalues $m,$ $n,$ $p$ and $q,$ respectively, which will be denoted by $R^{m,n,p,q}.$ Again, we have step-up and step-down equations, which we will not write explicitly, and $R^{0,0,0,0}$ is the unit trace eigenstate whose indices cannot be decreased. Its explicit form is 

\begin{equation}\label{modoescuro}
R^{0,0,0,0} = R^{0,0}_a R^{0,0}_b \, ,
\end{equation}
where $R^{0,0}_{\sigma} = \vert 0 \rangle_{\sigma} \hskip 0.1cm _{\sigma}\langle 0 \vert ,$ with $\sigma = a,b,$ is the eigenstate of lowest eigenvalue for a single mode, according to Eq.~(\ref{R00}). The remaining eigenstates are obtained by stepping $R^{0,0,0,0}$ up, 

\begin{equation}
R^{m,n,p,q} =  \frac{{\cal Q}^q_+ }{ \sqrt{q!}} \frac{{\cal P}^p_+ }{ \sqrt{p!}} \frac{{\cal N}^n_+ }{ \sqrt{n!}} \frac{{\cal M}^m_+ }{ \sqrt{m!}} R^{0,0,0,0} \, ,
\end{equation}
and their explicit forms will be

\begin{equation}\label{Rmnpq}
R^{m,n,p,q} = \sum_{k=0}^{m+n} \sum_{l=0}^{p+q} D_k^{m,n} D_l^{p,q \hskip 0.1cm *} R^{k,l}_a R^{m+n-k,p+q-l}_b \, ,
\end{equation}
where

\begin{eqnarray}\label{D}
D_k^{m,n} &=& \sqrt{\frac{k! (m+n-k)! }{ m! n!}} \sum_{j={\rm max} (0,k-n)}^{{\rm min} (m,k)} \frac{m! }{ j! (m-j)!} \nonumber \\
&\times& \frac{n! }{ (k-j)!(n-k+j)!} r_+^j s_+^{m-j} r_-^{k-j} s_-^{n-k+j}
\end{eqnarray}
and $R^{m,n}_{\sigma},$ with $\sigma = a,b,$ are the eigenstates for the case of a single mode given in Eq.~(\ref{autoestado}). Thus, the solution of the master equation for the case of two coupled modes in contact with a zero-temperature reservoir is

\begin{eqnarray}\label{solucaoab}
\rho(t) &=& \sum_{m=0}^{\infty}\sum_{n=0}^{\infty}\sum_{p=0}^{\infty}\sum_{q=0}^{\infty} C^{m,n,p,q} R^{m,n,p,q} \nonumber \\
&\times& e^{\displaystyle{(m \lambda_+  + n \lambda_- + p \lambda_+^* + q \lambda_-^*)t}}\, ,
\end{eqnarray}
where the expansion coefficients are

\begin{equation}\label{Cmnpq}
C^{m,n,p,q} = {\rm tr} \biggl(\frac{{\cal Q}_-^q }{ \sqrt{q!}} \frac{{\cal P}_-^p }{ \sqrt{p!}} \frac{{\cal N}_-^n }{ \sqrt{n!}} \frac{{\cal M}_-^m }{ \sqrt{m!}} \rho(0) \biggr)\, .
\end{equation}
By defining the operators $\chi_{\pm} = u_{\pm}a + v_{\pm}b,$ we may rewrite these coefficients in a simpler form,

\begin{equation}\label{Cmnpqexplicito}
C^{m,n,p,q} =\frac{1  }{ \sqrt{m! n! p! q!}} \hskip 0.1cm {\rm tr} (  \chi_-^n \chi_+^m  \hskip 0.1cm \rho(0) \hskip 0.1cm \chi_+^{\dagger \hskip 0.1cm p} \chi_-^{\dagger \hskip 0.1cm q} )\, .
\end{equation}

As example, we may consider the case in which both modes are in a coherent state

\begin{equation}\label{R0ab}
\rho(0) = \vert \alpha \rangle_{a} \hskip 0.1cm _{a}\langle \alpha \vert \hskip 0.1cm \vert \beta \rangle_{b} \hskip 0.1cm _{b}\langle \beta \vert\, .
\end{equation}
In this case, the expansion coefficients will be given by

\begin{eqnarray}\label{Cab}
C^{m,n,p,q}_{\alpha, \beta} &=& \frac{(u_+ \alpha + v_+ \beta)^m }{ \sqrt{m!}} \hskip 0.1cm \frac{(u_- \alpha + v_- \beta)^n }{ \sqrt{n!}}  \nonumber \\
&\times& \frac{(u_+^* \alpha^* + v_+^* \beta^*)^p }{ \sqrt{p!}} \hskip 0.1cm  \frac{(u_-^* \alpha^* + v_-^* \beta^*)^q }{ \sqrt{q!}}
\end{eqnarray}
and the solution of the master equation is

\begin{equation}\label{Rtab}
\rho(t) = \vert \alpha(t) \rangle_{a} \hskip 0.1cm _{a}\langle \alpha(t) \vert \hskip 0.1cm \vert \beta(t) \rangle_{b} \hskip 0.1cm _{b}\langle \beta(t) \vert \, ,
\end{equation}
where

\begin{eqnarray}\label{alfat}
\alpha(t) &=& \alpha F(t) + \beta G(t)\, , \nonumber \\
\beta(t)  &=& \alpha H(t) + \beta I(t)\, , 
\end{eqnarray}
being

\begin{eqnarray}\label{funcoes}
\pmatrix{F(t) & G(t) \cr H(t) & I(t) \cr} &=& \pmatrix{r_+ & r_- \cr s_+ & s_- \cr} \nonumber \\
&\times& \pmatrix{e^{\displaystyle{\lambda_+t}} & 0 \cr 0 & e^{\displaystyle{\lambda_-t}} \cr} \pmatrix{u_+ & v_+ \cr u_- & v_- \cr}\, .
\end{eqnarray}
With the coefficients (\ref{coeficientes}), we will obtain the explicit form of these functions

\begin{eqnarray}\label{funcoesexplicitas}
F(t) &=& \biggl(cos (\Delta t) - i \frac{S }{ \Delta} sen (\Delta t) \biggr)e^{\displaystyle{-iRt}}\, , \nonumber \\
G(t) &=& -i \frac{U }{ \Delta} sen(\Delta t)e^{\displaystyle{-iRt}}\, , \nonumber \\
H(t) &=& -i \frac{V }{ \Delta} sen(\Delta t)e^{\displaystyle{-iRt}}\, , \nonumber \\
I(t) &=& \biggl(cos (\Delta t) + i \frac{S }{ \Delta} sen (\Delta t) \biggr)e^{\displaystyle{-iRt}}\, .
\end{eqnarray}
We see that, in this example, the two modes are always separable.

\section{Two coupled modes in contact with a single thermal reservoir}

We want now to include temperature in the system formed by two coupled modes in contact with a single thermal reservoir. It only will affect the Lindblad superoperator that will be now \cite {sonia}

\begin{eqnarray}\label{Ldoismodos}
{\cal L} &=& (\bar n + 1)\frac{\gamma_a }{ 2} (2 a..a^{\dagger} - a^{\dagger} a. - .a^{\dagger} a)\nonumber \\ 
&+& \bar n \frac{\gamma_a }{ 2} (2 a^{\dagger}..a - a a^{\dagger}. -.a a^{\dagger}) \nonumber \\  
&+& (\bar n + 1)\frac{\gamma_b }{ 2} (2 b..b^{\dagger} - b^{\dagger} b. - .b^{\dagger} b) \nonumber \\
&+& \bar n \frac{\gamma_b }{ 2} (2 b^{\dagger}..b - b b^{\dagger}. -.b b^{\dagger}) \nonumber \\
&+& (\bar n + 1)\frac{\gamma_c }{ 2} (2 b..a^{\dagger} - a^{\dagger} b. - .a^{\dagger} b) \nonumber \\
&+& \bar n \frac{\gamma_c }{ 2} (2 a^{\dagger}..b - b a^{\dagger}. -.b a^{\dagger}) \nonumber \\ 
&+& (\bar n + 1)\frac{\gamma_c^* }{ 2} (2 a..b^{\dagger} - b^{\dagger} a. - .b^{\dagger} a) \nonumber \\
&+& \bar n \frac{\gamma_c^* }{ 2} (2 b^{\dagger}..a - a b^{\dagger}. -.a b^{\dagger}) \, ,
\end{eqnarray}
where $\vert \gamma_c \vert = \sqrt{\gamma_a \gamma_b}.$ The case of two reservoirs at the same temperature may be obtained taking $\gamma_c = 0.$ 

The step-up superoperators are now given by

\begin{eqnarray}\label{levantamentoabT}
\widetilde{\cal M}_+ &=& r_+ \widetilde{\cal M}_{+}^{a} + s_+  \widetilde{\cal M}_{+}^{b} \, , \nonumber \\
\widetilde{\cal N}_+ &=& r_-  \widetilde{\cal M}_+^{a} + s_-  \widetilde{\cal M}_+^{b} \, , \nonumber \\
\widetilde{\cal P}_+ &=& r_+^*  \widetilde{\cal N}_+^{a} + s_+^*  \widetilde{\cal N}_+^{b}\, , \nonumber \\
\widetilde{\cal Q}_+ &=& r_-^*  \widetilde{\cal N}_+^{a} + s_-^*  \widetilde{\cal N}_+^{b} \, ,
\end{eqnarray}
whereas the step-down superoperators are

\begin{eqnarray}\label{levantamentoabT}
\widetilde{\cal M}_- &=& u_+  \widetilde{\cal M}_-^{a} + v_+  \widetilde{\cal M}_-^{b}\, , \nonumber \\
\widetilde{\cal N}_- &=& u_-  \widetilde{\cal M}_-^{a} + v_-  \widetilde{\cal M}_-^{b}\, , \nonumber \\
\widetilde{\cal P}_- &=& u_+^*  \widetilde{\cal N}_-^{a} + v_+^*  \widetilde{\cal N}_-^{b}\, , \nonumber \\
\widetilde{\cal Q}_- &=& u_-^* \widetilde{\cal N}_-^{a} + v_-^*  \widetilde{\cal N}_-^{b}\, ,
\end{eqnarray}
where the superoperators with indices $a,b$ refer to those for the case of a single thermal mode of the form (\ref{superoperadoresT}) (for example, $\widetilde{\cal M}_-^{a} = - \bar n .a + (\bar n + 1)a.$ and $\widetilde{\cal M}_-^{b} = - \bar n .b +(\bar n + 1)b.).$ The steady state for the case of two coupled thermal oscillators is factorable into the steady states of each mode of the form (\ref{R00T}),

\begin{equation}\label{R00abT}
R^{0,0,0,0}(\bar n) = R^{0,0}_a(\bar n) R^{0,0}_b(\bar n)\, ,
\end{equation}
indicating that the steady state is the one in which each oscillator is in thermal equilibrium with the reservoir. The remaining eigenstates are obtained by stepping the steady state up, 

\begin{equation}
R^{m,n,p,q}(\bar n) =  \frac{\widetilde{\cal Q}^q_+ }{ \sqrt{q!}} \frac{\widetilde{\cal P}^p_+ }{ \sqrt{p!}} \frac{\widetilde{\cal N}^n_+ }{ \sqrt{n!}} \frac{\widetilde{\cal M}^m_+ }{ \sqrt{m!}} R^{0,0,0,0}(\bar n) \, ,
\end{equation}
resulting in

\begin{eqnarray}\label{Rmnpq}
R^{m,n,p,q}(\bar n) &=& \sum_{k=0}^{m+n} \sum_{l=0}^{p+q} D_k^{m,n} D_l^{p,q \hskip 0.1cm *} \nonumber \\
&\times& R^{k,l}_a (\bar n) R^{m+n-k,p+q-l}_b(\bar n)\, ,
\end{eqnarray}
where $D_k^{m,n}$ are given in Eq.~(\ref{D}) and $R^{m,n}_{\sigma} (\bar n),$ with $\sigma = a,b,$ are the eigenstates of a single thermal mode given in Eq.~(\ref{autoestadoT}).

The expansion of the density operator into the eigenstates will be similar to Eq.~(\ref{solucaoab}),

\begin{eqnarray}\label{solucaoabT}
\rho_T(t) &=& \sum_{m=0}^{\infty}\sum_{n=0}^{\infty}\sum_{p=0}^{\infty}\sum_{q=0}^{\infty} C^{m,n,p,q}(\bar n) R^{m,n,p,q}(\bar n) \nonumber \\
&\times& e^{\displaystyle{(m \lambda_+  + n \lambda_- + p \lambda_+^* + q \lambda_-^*)t}}\, ,
\end{eqnarray}
where the expansion coefficients are given by

\begin{equation}\label{CmnpqT}
C^{m,n,p,q}(\bar n) = {\rm tr} \biggl(\frac{\widetilde{\cal Q}_-^{\hskip 0.1cm q} }{ \sqrt{q!}} \frac{\widetilde{\cal P}_-^{\hskip 0.1cm p} }{ \sqrt{p!}} \frac{\widetilde{\cal N}_-^{\hskip 0.1cm n} }{ \sqrt{n!}} \frac{\widetilde{\cal M}_-^{\hskip 0.1cm m} }{ \sqrt{m!}} \rho(0) \biggr)\, .
\end{equation}
                                                                                                                                           
\section{Conclusions}

To summarize, we have developed a simple algebraic method to obtain the solution of master equations with quadratic Liouville superoperators for finite temperatures and any initial conditions. This method consists in the expansion of the initial density operator into the eigenstates of the Liouvillian, allowing its evolution in a simple form. For obtaining these eigenstates we have used ladder superoperators, which are determined through commutation relations with the Liouvillian. We have found the steady state and the expansion coefficients by using step-down superoperators and, the remaining normal modes, by stepping the steady state up.  
 
\acknowledgments

We want to thank N. Zagury for very helpful discussions. This work was supported by the
Brazilian agency: CNPq.

\appendix

\section{Completeness of the basis formed by eigenstates of one non-thermal mode}

We want to demonstrate that the eigenstates $R^{m,n}$ with positive integer indices form a complete set, i.e., they can expand any target operator $\rho.$ For this demonstration we will use a straightforward calculation, i.e., we will use the explicit form of these eigenstates and of the expansion coefficients and suitably sum the series. 

The expansion coefficients can be developed by performing the trace of the target operator $\rho$ in the basis of the oscillator eigenstates and applying the operators $a$ and $a^{\dagger}$ presents in the step-down superoperators (\ref{superoperadores})

\begin{eqnarray}\label{coef}
{\rm tr} \biggl(\frac{{\cal N}_-^n }{ \sqrt{n!}} \frac{{\cal M}_-^m }{ \sqrt{m!}} \hskip 0.1cm \rho \biggr) = \sum_{l=0}^{\infty} \frac{1 }{ l!} \sqrt{\frac{(m+l)! }{ m!}\frac{(n+l)! }{ n!}} \nonumber \\
\times  \langle m+l \vert \rho \vert n+l \rangle \, .
\end{eqnarray}
Using the explicit form of the eigenstates $R^{m,n},$ given in Eq.~(\ref{autoestado}), we can sum the series

\begin{eqnarray}\label{serie}
& &\sum_{m=0}^{\infty} \sum_{n=0}^{\infty} {\rm tr} \biggl(\frac{{\cal N}_-^n }{ \sqrt{n!}} \frac{{\cal M}_-^m }{ \sqrt{m!}} \hskip 0.1cm \rho \biggr) R^{m,n} \nonumber \\
&=&  \sum_{k=0}^{\infty} \sum_{m=k}^{\infty} \sum_{n=k}^{\infty} \sum_{l=0}^{\infty} \frac{(-1)^k }{ k! l!} \sqrt{\frac{(m+l)! }{ (m-k)!}\frac{(n+l)! }{ (n-k)!}} \nonumber \\
&\times& \langle m+l \vert \rho \vert n+l \rangle \hskip 0.1cm \vert m-k \rangle \langle n-k \vert \, ,
\end{eqnarray}
where we have suitably reordered the summations. By changing the sum indices $m \rightarrow m+k,$ $n \rightarrow n+k$ and $l \rightarrow l-k$ and reordering the summations, we get

\begin{eqnarray}\label{trocadesoma}
& &\sum_{m=0}^{\infty} \sum_{n=0}^{\infty} {\rm tr} \biggl(\frac{{\cal N}_-^n }{ \sqrt{n!}} \frac{{\cal M}_-^m }{ \sqrt{m!}} \hskip 0.1cm \rho \biggr) R^{m,n} \nonumber \\
&=&  \sum_{m=0}^{\infty} \sum_{n=0}^{\infty} \sum_{l=0}^{\infty} \sum_{k=0}^{l} \frac{(-1)^k }{ k! (l-k)!} \sqrt{\frac{(m+l)! }{ m!}\frac{(n+l)! }{ n!}} \nonumber \\
&\times& \langle m+l \vert \rho \vert n+l \rangle \hskip 0.1cm \vert m \rangle \langle n \vert \, .
\end{eqnarray}
The summation over $k$ is easily done resulting in $\delta_{l,0},$ which eliminates the summation over $l.$ It is easy to recognize that the resulting expression is exactly the expansion of the operator $\rho$ into the basis of the oscillator eigenstates. This allows us to obtain

\begin{equation}\label{completeza}
\sum_{m=0}^{\infty} \sum_{n=0}^{\infty} {\rm tr} \biggl(\frac{{\cal N}_-^n }{ \sqrt{n!}} \frac{{\cal M}_-^m }{ \sqrt{m!}} \hskip 0.1cm \rho \biggr) R^{m,n} =   \rho \, .
\end{equation}
This last expression shows that the set of the eigenstates $R^{m,n}$ with positive integer indices is complete.

In this proof, it was essential to reorder the summations, but this is not always possible because it requires certain conditions. Then the above demonstration will not be always valid. As criterion to determine the validity of (\ref{completeza}), we have that this equality will occur if and only if each matrix element of the expansion defined in its left-hand side converges. A simple and concrete example of this fact appears in the case of a thermal distribution 

\begin{equation}\label{Rx}
\rho(\bar n) = \sum_{n=0}^{\infty} \frac{\bar n^k }{ (\bar n+1)^{k+1}} \vert k \rangle \langle k \vert \, . 
\end{equation}
Its expansion into the eigenstates will give us

\begin{eqnarray}
\sum_{m=0}^{\infty}\sum_{n=0}^{\infty} {\rm tr} \biggl(\frac{{\cal N}_-^n }{ \sqrt{n!}}  \frac{{\cal M}_-^m }{ \sqrt{m!}}\hskip 0.1cm \rho(\bar n) \biggr) \hskip 0.1cm R^{m,n} \nonumber \\
= \sum_{k=0}^{\infty} \sum_{n=k}^{\infty} (-1)^{n-k} \frac{n! }{ k! (n-k)!} \bar n^n \vert k \rangle \langle k \vert \, .
\end{eqnarray}
This expansion correctly converges for (\ref{Rx}) when $\bar n < 1,$ due to identity (\ref{identidade}), but diverges when $\bar n \geq 1.$

The completeness for the remaining cases can be also demonstrated, however it is more arduous and therefore will be omitted.


\begin{thebibliography}{99}
\bibitem {identity} This identity is valid for $\vert x \vert <1$ and is easily obtained by differentiating $k$ times the expansion of the function $(1+x)^{-1}$ into power series.
\end{thebibliography}
\end{document}